# Introducing PathQuery, Google's Graph Query Language

Jesse Weaver
Google
jrweave@google.com

Eric Paniagua
Google
epaniagua@google.com

Tushar Agarwal
Google
agarwaltushar@google.com

Nicholas Guy
Google
nguy@google.com

Alexandre Mattos
Google
amattos@google.com

## ABSTRACT

We introduce PathQuery, a graph query language developed to scale with Google's query and data volumes as well as its internal developer community. PathQuery supports flexible and declarative semantics. We have found that this enables query developers to think in a naturally "graphy" design space and to avoid the additional cognitive effort of coordinating numerous joins and subqueries often required to express an equivalent query in a relational space.

Despite its traversal-oriented syntactic style, PathQuery has a foundation on a custom variant of relational algebra – the exposition of which we presently defer – allowing for the application of both common and novel optimizations.

We believe that PathQuery has withstood a "test of time" at Google, under both large scale and low latency requirements. We thus share herein a language design that admits a rigorous declarative semantics, has scaled well in practice, and provides a natural syntax for graph traversals while also admitting complex graph patterns.

## 1 INTRODUCTION

Graph data come to the fore in several real-world contexts [7, 8, 10, 11], and a number of graph data models and graph query languages have been proposed [1, 5, 6, 9, 12] and classified in recent surveys [3, 4, 13]. In this paper, we share some of Google's work on graph querying, specifically the Path-Query language. Using the terminology of [3], PathQuery is *homomorphism-based* with *bag* semantics, featuring *navigation*-oriented syntax. Additionally, it has a formal algebraic foundation.

We review the motivation for our work in Section 2 before describing PathQuery's data and execution models as well as the language itself in Section 3. Section 4 briefly discusses how PathQuery is used within Google, and we look at where PathQuery fits among some popular, graph query languages in Section 5. We then conclude with a brief summary in Section 6.

## 2 MOTIVATION

Google announced its Knowledge Graph (KG) in 2012: "The Knowledge Graph enables you to search for things, people or places that Google knows about ... and instantly get information that's relevant to your query" [11]. To facilitate access to and complex querying of the KG, a graph query language seemed desirable, if not necessary. At that time, there appeared to be roughly two categories of graph query languages: (1) those similar to SQL, and (2) proprietary languages. Google needed a graph query language that could scale to its large number of developers without requiring deep expertise. Thus, to be intuitive, the graph query language needed to be naturally "graphy", which in our opinion ruled out the languages that were similar to SQL. Such languages – while having basic graph pattern matching mechanisms – ultimately require pivoting a developer's mindset into a relational space, instead of a "graphy" space, when attempting any non-trivial query. Additionally, proprietary languages were *proprietary*, offering little or no formal foundation for semantics, much less any well-defined opportunities for query optimization.

Thus, Google endeavored to create PathQuery, a "graphy" graph query language that has scaled in development, performance, and growth of Google's KG. Although not discussed in detail in this particular paper, PathQuery is grounded in a custom variant of relational algebra that allows for both common and novel optimizations. However, this grounding is not significantly reflected in the query language, thus keeping the language as graph-oriented as possible and lessening the cognitive effort of developers. PathQuery has become the interface to the KG for many Google products – including Search, Maps, and Assistant – where latency and efficiency requirements are of paramount consideration. Thus, we believe the language has withstood significant, practical challenges over time and would be worth sharing with the community at large. We write this paper to introduce the PathQuery language and to compare and contrast it with existing graph query languages (see Section 5). We begin with an introduction to the PathQuery language in the following section.

## 3 THE PATHQUERY LANGUAGE

PathQuery can be conceptualized as a declarative dataflow language. Data logically flow from the beginning toward the end of the query, roughly left-to-right.[1]

It is also important at this point to note that "PathQuery" is the name of our language largely for historical reasons. As will be shown further in this paper, the language has a strong navigational feel, **but has no particular relationship to the general concept of** *path queries*, e.g., as defined in [3].

We briefly discuss the model in Section 3.1 before exploring the PathQuery language through examples in Section 3.2. We then proceed to discuss language constructs more formally in Section 3.3.

---

[1]The true order of execution is decided by a query planner.

### 3.1 Model

*3.1.1 Data Types.* PathQuery supports a number of data types, listed in Table 1, and every value has one of those types. Most of these are common types, but two are more idiosyncratic: `Id` and Record. `Id` values identify entities in the graph and are not equivalent to string representations of such identifiers. Record values are tree-like data structures with named fields, each mapped to one or more values, any of which may further be a record. They are often helpful for structuring the results of a query, associating values during query processing, and propagating groupings of data from earlier parts of the query to later parts.

PathQuery requires an ordering on values that has certain properties. The ordering must be nearly total, though some details are implementation-defined. This ordering is the default ordering used by `Max`, `Min`, `Sort`, and `Top`, which are discussed below. It is worth noting that record values themselves are not comparable. They may, if explicitly indicated, be sorted by the values of their fields.

*3.1.2 Data Model.* PathQuery was originally designed with only Google's KG in mind. The KG is effectively a set of *triples*, each of which consists of three parts: a *subject*, a *predicate*, and an *object*. From PathQuery's perspective, KG subjects are always `Id`s, whereas objects are any non-Record value. The predicate is basically a string, although PathQuery does not view predicates as values.

More generally, building on terminology from [3], Path-Query evaluates over any *edge-labelled graph*. KG triples can be viewed as directed edges between nodes, where only nodes identified with `Id`s are allowed to have outgoing edges. The only requirement is that node identifiers must belong to one of PathQuery's non-Record data types.

*3.1.3 Execution Model.* The PathQuery language implicitly assumes that there exists a single target graph over which queries run, so in the rest of the paper, we mostly refrain from mentioning the target graph explicitly.

The first concept we need to establish is that of a *collection*. A collection is a multiset of values, whose permissible types are drawn from Table 1.

Syntactically, PathQuery consists of *paths*, some of which are fundamental, others of which are compositions of other paths. Semantically, a path is a function from collections to collections. We discuss paths as having *inputs* and *outputs*, where "the inputs" is a collection *coming in* to the path, and "the outputs" is a collection *going out* of the path. A whole query is itself a path, but with a semantic exception that the first path(s) in the query determine(s) the *root values*, i.e., the values from which evaluation of the query logically begins.

The result of a query is a mapping from the root values to the output values at the end of the query.[2] This overall result can be thought of as a single Record value if desired, although such is not specified by the language itself; it could just as easily be a collection of pairs.

### 3.2 PathQuery by Example

We begin by building an extended example to develop an intuitive understanding of the PathQuery language. Suppose we are curious about attractions to visit in New York City on a Friday. More specifically, we want to find information about museums and theme parks in the area. Our strategy will be to find all museums and theme parks and to continue filtering until we are satisfied with the results. We start by finding all museums and theme parks.

**attractions_v1.pq**

```
1 @entities
2   .[/type == (Id('/museum'), Id('/theme_park'))]
3   ./name
```

We consider the collection of all entities, `@entities`, on line 1. The bracketed construct on line 2 is a *where clause*. A *where clause* outputs only those inputs for which the path wrapped in brackets has at least one output value that is true.[3] Here, we are checking – for each entity – whether any outgoing /type edges point to either the `Id('/museum')` or `Id('/theme_park')` nodes.[4] The next element to note is the '.' (Dot) at the start of lines 2 and 3. *Dot* denotes composition, i.e., that outputs from the path on its left (such as `@entities` on line 1) are passed on as inputs to the path on its right (such as the where clause on line 2). On line 3, another Dot passes on the outputs of the where clause as the inputs to the predicate /name. A predicate used in this fashion effectively traverses the outgoing /name edges from the nodes identified by the inputs, and outputs values identifying any nodes that are reached from the traversal. In short, predicates are the syntactic structure for edge traversal.

The following is an example result.[5]

**Sample Output for attractions_v1.pq**

```
Id('/z/38dwfnb8'):
  Text('Museum of Modern Art', 'en')
Id('/z/38dwfnb8'):
  Text('Museo de Arte Moderno', 'es')
Id('/z/hklbdfap'): Text('Family Park', 'en')
Id('/z/hklbdfap'): Text('Parc familial', 'fr')
Id('/z/389_8934r'):
  Text('Natural History Museum', 'en')
...

(total results: 495258)
```

---

[2]Output values for non-trivial queries are often records in practice.

[3]Every value other than the `Bool` value `false` is considered to be effectively true.

[4]Entity *types* in the KG are themselves entities and so are identified by values of the `Id` data type, as in this example. However, PathQuery has no such restriction.

[5]The `/z/...` format of `Id` values is contrived; PathQuery places no restriction on what `Id`s look like.

**Table 1: Data Types of PathQuery**

| Type | Description | Examples |
|------|-------------|----------|
| `Bool` | True or false | `true`<br>`false` |
| `DateTime` | An ISO 8601 DateTime with no subseconds | `DateTime('2019-10-31T08:00:00Z')`<br>`DateTime('2019-10-31T08:00:00')`<br>`DateTime('2019-10-31T08:00:00+05:00')`<br>`DateTime('2019-10-31')`<br>`DateTime('T08:00')` |
| `Double` | A double-precision floating point value | `5.1`<br>`-0.01e-15`<br>`Double('5')`<br>`Double('inf')`<br>`Double('-inf')` |
| `Duration` | An ISO 8601 Duration without years, months, or weeks | `Duration('PT1H')`<br>`Duration('P30D')`<br>`Duration('PT1M30S')` |
| `Id` | A special identifier for entities in the KG | `Id('/z/14znzk')` |
| `Int` | A 64-bit 2's-complement integer | `5`<br>`-2000`<br>`Int('7')` |
| `Record` | A record composed of literals | `{ f: 5 }` |
| `String` | A string value | `'single quotes'`<br>`"double quotes"` |
| `Text` | A string value with an associated language tag | `Text('hello world', 'en')` |

As indicated earlier, the result is a mapping from the root values to the output values. In this case, the root values are entity `Id`s, and they are mapped to names of museums and theme parks. Note that a single root value may correspond to multiple output values.

Continuing with our example, there are clearly too many results to sift through manually, and the results are reported in multiple languages. Suppose we want to limit the results to a single language and make them more informative.

**attractions_v2.pq**

```
1 import 'events.pq'
2
3 @entities
4 ./[/type == (Id('/museum'), Id('/theme_park'))]
5 .{
6     id: ?cur
7     require name: /name.[TextLang() == 'en']
8     @merge: events::GetInfo()
9 }
```

Our new addition (attractions_v2.pq:5-9) constructs a *record* – one for each attraction – to provide additional information in a structured format. Each of lines 6-8 denotes a *field* of the record. The `id` field has as its value `?cur`, denoting the "current value," which in this case is the input from the Dot for which a record is being created, on line 5. The previously retrieved names are now captured in the `name` field, although this time, we have used a where clause to restrict the values to only English results. Note that the `name` field is annotated with the `require` keyword, indicating that it must have at least one value in order for the record to be constructed. Without the `require` keyword, a field is completely optional. Finally, `@merge` is a special field name whose value must be a record and which merges the fields of that record into the one with the `@merge`. In this case, the value to be merged is computed by a function `GetInfo` from the `events` module (not explicitly defined in any example herein) imported on line 1. Note that if the *function call* `events::GetInfo()` produces no output, a record will still be created because the `@merge` field is *not* marked with the `require` keyword.

Below is a new example result. Note that the `open_hours` field is present because it was merged into the result by the use of `@merge`.

**Sample Output for attractions_v2.pq**

```
Id('/z/38dwfnb8'): {
  id: Id('/z/38dwfnb8')
  name: Text('Museum of Modern Art', 'en')
  open_hours: {
    days: 'MTWR'
    start: DateTime('T09:00')
    end: DateTime('T18:00')
  }
  open_hours: {
    days: 'SU'
    start: DateTime('T10:00')
    end: DateTime('T15:00')
  }
}
...
(total results: 483425)
```

Continuing with our example, suppose we notice a few things in our new results:

- The Museum of Modern Art is closed on Fridays, but we are only interested in attractions that are open on Fridays.
- 11,833 results were removed because they did not have /name triples with the 'en' language tag.
- We notice that a number of attractions are closed permanently.
- We notice many attractions that are nowhere near New York City (e.g., the Louvre).
- We would like to see information about the rides and exhibits.
- There are still too many results for us to sift through manually.

Continuing, we make a third pass at composing our query, in which we restrict ourselves to the 25 top results based on distance from New York City.

**attractions_v3.pq**

```
1  import 'events.pq'
2  import 'spacetime.pq' into st
3
4  def EnglishName() {
5    /name.[TextLang() == 'en']
6  }
7
8  // This function is defined in a library linked
9  // into the executable that will run our query.
10 external def expr Distance(?geometry,
11                            ?location_of_interest,
12                            ?radius_km)
13   implemented_by 'Distance'
14
15 def ComputeDistance() {
16   Distance(st::FeatureGeometry(),
17            ?params->location_of_interest,
18            ?params->radius_km)
19 }
```

```
20
21 Sort(Top(
22   @entities.{
23     [/type == (Id('/museum'),
24                Id('/theme_park'))];
25     require((/exhibit, /ride));
26     require(EnglishName());
27     prohibit(/permanently_closed.[?cur]);
28     [st::IsOpenOnDayOfWeek(
29       /opening_hours,
30       ?params->day_of_week)];
31     ComputeDistance().bind(?distance)
32   },
33   ?params->max_num_results,
34   ?distance, ?root)
35 .?root
36 .{
37   id: ?cur
38   name: EnglishName()
39   @merge: events::GetInfo()
40   distance: ?distance
41   mean_price: {
42     require def $prices classify {
43       [/type == Id('/museum')]: {
44         /exhibit
45       }
46       [/type == Id('/theme_park')]: {
47         /ride
48       }
49     }./price;
50     Sum($prices) / Count($prices)
51   }
52 },
53 ?cur->distance, ?cur->id)
```

Our third pass (attractions_v3.pq) introduces a wealth of new language features.

- The query begins on line 21 with a call to Sort, indicating that the results will be sorted. The first argument to Sort is the significant remainder of the query on lines 21-52. Line 53 specifies the sort order, which is ascending on distance field values, and then as a tie-breaker, on id field values.
  - This is the first time we have seen the *field access* operator (->). The field access operator simply outputs the values stored in the named field.
- Returning to line 21, we see a call to Top, indicating that we expect to get some top-$k$ results. The first argument to Top is on lines 22-32. Line 33 specifies the number of results desired. Line 34 specifies the ordering criteria, which is first by distance, and then by the root value (see below) as a tie-breaker.
  - Line 33 is the first time we have used the special ?params variable. When executing a query, a record value may be provided *alongside* the query and made available via ?params, which is globally visible. This provides a way to parameterize the query.

– Line 34 is the first time we have used the special `?root` variable. This variable is globally visible and is always assigned to the root value that led to this point in the query.

### `?params` for attractions__v3.pq

```
{
  location_of_interest: 'New York, NY, USA'
  radius_km: 10.0
  day_of_week: 'F'
  max_num_results: 25
}
```

- On line 22, the root values are established by beginning with `@entities`, which outputs `Ids` for all entities in the graph. The outputs of `@entities` become the inputs to a *block*, syntactically enclosed in curly braces.
  - A block allows for traversing across "tangents" from a single point. That is, the inputs to the element paths, separated by semicolons, on lines 23-31 are the same as the inputs to the block. Intuitively, each element is "traversed" to the end, at which point traversal "restarts" at the input value of the block, and begins with the next element. Earlier elements can have a filtering/cardinality effect on later elements. The outputs of the last element become the outputs for the block.
- Line 23 filters entities down to museums and theme parks.
- Line 25 shows the `require` path[6] which is a pure existence check. For each input value to the `require` path, if (/exhibit, /ride) outputs *any* values for that input value, then the `require` path outputs the input value; otherwise, the `require` path outputs nothing for that input value.
  - (/exhibit, /ride) is the first time we have seen a *tuple* path, which is denoted by parentheses. A *tuple* expresses a divergence and convergence of paths. Each input value to the tuple becomes an input value to both /exhibit and /ride, and all the output values from those two predicates become the collective outputs of the tuple for that input value.
- On line 26, we filter out entities that do not have an English name. Note that we do not keep the name at this point. Because we are still within the `Top(...)` argument, we are focused on limiting the selection of the top 25 entities before we look up values, which we do *after* the `Top` call.[7]
- On line 27, we see the `prohibit` path for the first time. The `prohibit` path is a pure *non*-existence check. For each input value to the `prohibit` path, if /permanently_closed.[?cur] outputs *any* values, then the `prohibit` path outputs nothing for that input value; otherwise, the `prohibit` path outputs the input value.

  – It is worth noting here that the `prohibit` path is filtering out entities that are known to be permanently closed. It allows for entities that are *not* permanently closed (i.e. /permanently_closed points to `false`) and for entities for which such status is unknown (i.e. there are *no* outgoing edges labelled /permanently_closed).
- On lines 28-30, we keep only entities that are open on the specified `?params`->day_of_week. We pass the output value of /opening_hours as the first argument to a function call to `st::IsOpenOnDayOfWeek`. Note that the function is in the `st` namespace, which is the namespace into which the `spacetime.pq` module was imported on line 2.
- On line 31, we call `ComputeDistance` to compute the distance of the entity to `?params`->location_of_interest (but only within a radius of `?params`->radius_km).
  – Note that in the definition of `ComputeDistance` (defined on lines 15-19), there is a call to another function named `Distance`, which is declared as *externally defined* on lines 10-13. An externally-defined function is a function implemented in some native, non-PathQuery language but which is callable from PathQuery.
  – On line 31, we also bind the output value of the call to ComputeDistance() to the variable ?distance. We do this because we want to save the value for multiple uses later. A `bind` path always outputs its input.
- On line 35, we have the first part of the query following the call to `Top`, and it is the invocation of the `?root` variable. When a variable is invoked, it simply replaces each of its input values with the value that is bound to the variable.[8]
- On lines 36-52, we build records. Note that at this point, we are limited to (at most) 25 input values, so we will build (at most) 25 records.
- The fields `id`, `name`, `@merge`, and `distance` are straightforwardly inferrable from previous examples and descriptions. The `mean_price` field shows some new features. It begins by opening a block and defining a collection `$prices`. Note the `require` keyword which indicates that this part of the query should not continue if `$prices` is empty. We then see a `classify` path which routes input values to appropriate cases. In the example, we see that inputs for museum entities will traverse outgoing /exhibit edges, while inputs for theme park entities will traverse outgoing /ride edges. Regardless of the case, the outputs then become the inputs to the /price predicate on line 49. Then on line 50, we use aggregates `Sum` and `Count` over the values in `$prices` to compute the average price.

---

[6]Not to be confused with the `require` keyword.

[7]Deferring value fetching and record creation until *after* the selection of entities is a common optimization in PathQuery.

[8]Note that the bound value *may* depend on the input value to its `bind` path. If the variable is undefined, it produces no output values.

## 3.3 Language Features

We will now survey PathQuery language features.

### 3.3.1 Basics.

`@entities`. For each input, output all entities in the graph.

*Predicates*. For each input, traverse all outgoing edges labelled with the predicate, and output all the nodes that were reached.

For traversing incoming edges, one need only prepend the predicate with the `!` character. That is, `/pred` denotes traversal of outgoing edges with the label "/pred", while `!/pred` denotes traversal of incoming edges with the label "/pred".

*Dot*. Dot is a pipe-like operator that passes the outputs from one path to the inputs of the next path. The Dot operator *may* reorder values.

`require`. For each input, a `require` path `require(P)` passes the input to P, and if P produces *any* outputs, then the `require` path outputs its input. Otherwise, the `require` path outputs nothing for that input.

`prohibit`. For each input, a `prohibit` path `prohibit(P)` passes the input to P, and if P produces *any* outputs, then the `prohibit` path outputs nothing. Otherwise, the `prohibit` path outputs the input.

`optional`. For each input, an `optional` path `optional(P)` passes the input to P, and if P produces *any* outputs, then those outputs become the outputs of the `optional` path for that input. Otherwise, the `optional` path outputs the input.

*Where Clause*. For each input, a where clause `[P]` passes the input to P, and if P produces *any non-false* outputs, then the where clause outputs the input. Otherwise, the where clause outputs nothing for that input.

*Tuples*. For each input, a tuple (P1, ..., Pn) passes the input to each of P1, ..., Pn, and the collective outputs of P1, ..., Pn become the outputs for the tuple for that input.

`classify`. For each input, a `classify` path `classify { C1: {P1} ... Cn: {Pn} }` effectively passes the input to each of C1, ..., Cn. The first one to have *any* outputs – say Ci – is "chosen", and the same input is passed to Pi. The outputs of Pi then become the outputs for the `classify` path for that input. If none of C1, ..., Cn has outputs, then an implicit default case is "chosen" that simply outputs the input. A `classify` path may explicitly specify the default case with the `else` keyword in place of Cn.

*Aggregates*. Basic aggregates reduce collections of values to at most a single value and include:

- For each input, `Count(P)` passes the input to P and then outputs a single number reflecting the number of outputs produced by P for that input, even if it is zero.
- For each input, `Min(P)` passes the input to P and then outputs a minimum value output by P for that input. The ordering used by `Min` may be customized in the same way that `Top`'s ordering was customized in the previous example.
- For each input, `Max(P)` passes the input to P and then outputs a maximum value output by P for that input. The ordering used by `Max` may be customized in the same way as for `Min`.
- For each input, `Sum(P)` passes the input to P and then outputs a single number that is the sum of the outputs of P for that input. If P has any non-numeric outputs, then `Sum(P)` has no output for the given input.

PathQuery also has aggregate-like operations (more similar to table-valued functions) that reduce collections of values to smaller collections. Examples include:

- For each input, `Dedup(P)` passes the input to P and then outputs the *unique* values output by P. Optional additional arguments may be specified to deduplicate by a custom notion of equivalence rather than by equality.
- For each input, `Slice(P, L, O)` passes the input to P and then outputs the first L values after the first O values output by P. Unless P provides a sorted output (like `Sort` does), `Slice` effectively selects an arbitrary collection of (at most) L values.

`Top`. For each input, `Top(P, K)` passes the input to P and then outputs the (at most) K smallest values output by P. As shown in previous examples, additional arguments can be specified to `Top` in order to customize ordering. For descending order, additional arguments may be preceded with '-', or for overall descending order, `Rtop` may be used instead.

`Sort`. For each input, `Sort(P)` passes the input to P and then outputs the outputs of P in ascending order. The same ability to customize ordering in `Top` applies to `Sort` as well.

*Blocks*. For each input, { P1; ...; Pn } passes the input to P1; then for each output of P1, the original input is passed to P2. This continues until Pn, at which point the outputs of Pn become the outputs of the block. In particular, if Pi produces no outputs for a given input, then that input does not propagate further through the block. Thus, blocks have the ability to filter as a special case of their ability to affect cardinality.

Blocks also provide scoping for anything defined using `def`, which includes functions, variables, and collections.

*Variables and Collections*. For each input, an invocation of the variable `?x` outputs the value bound to it. If `?x` is defined using the `def` keyword, the bound value is independent of the input value. If `?x` is defined with a `bind` path, the bound value may depend on the input. If `?x` is undefined, it outputs nothing.

For each input, an invocation of the collection `$c` outputs the values contained in it, which will be nothing if the collection is empty. Since collections cannot be defined using a `bind` path, the output of an invocation does not depend on the input.

Note that variables may be defined in two ways: with the `def` keyword in a block, and using a `bind` path. When defined with `def`, the variable is scoped to the innermost curly braces (i.e. roughly block scoped); when defined with `bind`, the variable is scoped to its innermost enclosing function, or it is query-scoped[9] if no such enclosing function exists. We allow any variable defined with `bind` to persist beyond its block because such a variable may be useful outside its block (such as `?distance` in attractions_v3.pq).

*3.3.2 Records.* Syntactically, records appear like blocks (i.e. curly braces) containing fields. Each field is a symbol (effectively a string, but not syntactically a string literal) followed by a path, the outputs of which become the values for the field.

*Record.* For each input, a record `{ f1: P1 ... fn: Pn }` passes the input to P1, ..., Pn, and for each Pi, the output values become the values for field f1. The instantiated record for that input becomes the output for the syntactic record.[10] If none of the fields have values, then the syntactic record outputs nothing. If any of the fields marked with `require` keyword have no value, then the syntactic record outputs nothing.

*Field Access.* For each input, a field access `P->f1->...->fn` passes the input to P. If an output of P is not a record, the field access outputs nothing for that input. If an output of P is a record, then the values in its f1 field are considered. Any non-record values of f1 are discarded. For the rest, the values in the f2 field are considered. It continues like this until fn, at which point the fn values become the outputs for the field access.

`@merge`. The special `@merge` field is used to build a larger record from smaller ones. For each input, a field `@merge: P` passes the input to P, and the fields of any records output by P become fields in the enclosing record.

*3.3.3 Functions.* PathQuery supports defining functions written directly in PathQuery. Combined with modules and imports (see Section 3.3.4) this provides a robust mechanism for abstraction, modularity, and reuse.

*Function Definitions.* A PathQuery function definition includes its name, zero or more variable or collection parameters, and a function body, as shown in attractions_v3.pq:15-19.

PathQuery also features a foreign function interface which allows it to call specially defined functions from natively-implemented libraries. The declaration of a function as externally defined is illustrated in attractions_v3.pq:10-13.

*Bounded Recursion.* PathQuery supports a restricted form of recursion where the number of recursive calls is bounded. The bound is provided in the function definition. Here is an example that looks up all the classifications for a given animal.

<div align="center">

**bounded.pq**

</div>

```
1  def EnglishName() {
2    /name.[TextLang() == 'en']
3  }
4  def NameAndRank() {
5    require name: EnglishName()
6    rank: /rank.EnglishName()
7  }
8  def base Classification() {
9    NameAndRank()
10 }
11 def recur<10> Classification() {
12   require @merge: NameAndRank()
13   under: /higher_classification
14     .Classification()
15 }
16 Id('/z/cat').Classification()
```

Note that two definitions of `Classification` are provided, the first annotated `base` and the second annotated `recur`<10>. The first defines the base case, used once the last allowed recursive call is reached, and the second defines the inductive case, used for all calls up to the last. The <10> in `recur`<10> says at most ten calls are allowed, including the final call to the base case.

*3.3.4 Modules and Imports.* PathQuery code can be shared across queries by factoring it into functions packaged into PathQuery modules. Queries and (other) modules that want to use those functions can access them by importing the module that defines them. We saw imports in our attractions_v2.pq and attractions_v3.pq examples above. Functions imported from modules are always namespaced to avoid naming collisions. A module is just a PathQuery file containing only function definitions or external function declarations.

*3.3.5 Parameterization.* Explicitly supporting query parameterization has a number of benefits. First, it obviates the need to construct queries at run time. Second, in doing so it improves system security by helping avoid injection-based attacks. Third, during debugging it becomes abundantly clear when logical changes to a query are made versus superficial changes.

We support parameterization in two ways:

`?params`. We saw `?params` in attractions_v3.pq. Explicitly, `?params` is a record-valued variable with global scope. Its value can be specified along with the query at runtime.

`@roots`. In our example queries so far, we have either started from the collection of all entities or we have hard-coded a specific entity from which to start. `@roots` allows a collection of entities to be provided alongside the query at runtime (much like `?params`), and these entities can then be used as starting points for the query. If no such entities are provided, `@roots` behaves the same as `@entities`.

---

[9] By "query-scoped", we mean that it is scoped to the "top level" (can be thought of as "main") of the query, but it is not truly *global* because it is not implicitly visible inside functions.

[10] To clarify, there is a difference between a record *value* and a *syntactic* record. A *syntactic* record is something written in the query that specifies the building of record *values*.

*3.3.6 Built-in Modules.* PathQuery comes with built-in standard modules, including but not limited to modules for time, math, strings, URLs, regexes, and geo support. Examples of extensive support include timezone support and geo functions like `AreaIntersects` for determining whether two geospatial regions overlap.

## 4 APPLICATION

At Google, PathQuery is used in a variety of contexts to support, e.g., Search, Maps, and Assistant.

- PathQuery is used for *serving online queries* over the KG, which demands high throughput and low latency over a very large graph.
- PathQuery is used for *offline queries* over the KG where the latency requirement is lifted, and the goal is to answer much broader queries for the purposes of extraction, validation, light analytics, etc.
- PathQuery is used for *in-memory evaluation of queries* as part of numerous other knowledge-related systems in Google, including ingestion pipelines and components of other serving systems.

## 5 COMPARISON WITH RELATED WORKS

Both storage and retrieval techniques for graph data have been extensively studied. Retrieval techniques range from specialized APIs to more general-purpose query languages. Fundamentally, graph query languages are domain-specific languages with the main purpose of interfacing with graph data. Modern graph query languages vary significantly in their semantics, style, and expressivity [3, 4, 13]. Here, we classify and briefly compare[12] PathQuery with SPARQL [1], Cypher [6], and Gremlin [9]. We roughly utilize the framework of [3] to present a summary in Table 2.

Query languages such as SPARQL (for RDF [2] data) and Cypher (for property graphs [3]) are declarative. They contain graph pattern matching mechanisms that can be composed using SQL-ish keywords. In contrast, Gremlin [9] is a functional-style language, where operations are chained together with a pipe-like '.' operator. PathQuery is much more in the vein of Gremlin's functional and "graphy" style rather than SPARQL's and Cypher's SQL-ish styles. However, PathQuery also maintains a declarative nature in that it is translated into a variant of relational algebra (although those details are admittedly not discussed herein). In this way, we feel that we have been able to harness the advantages of both kinds of languages: the declarative nature achieved by SQL-ish languages and the natively "graphy" feel of functional-style languages.

## 5.1 Complex Patterns

One of the dimensions of comparison in [3] is the ability to express complex patterns.[13] We feel this is a particularly important characteristic of a graph query language since any non-trivial query will need features beyond matching of basic graph patterns.

In this section, we write a query in four languages – including PathQuery – and provide a qualitative analysis. Building on previous examples, the query we desire to express is: "find all museum and theme park attractions that have a name and which are not known to be permanently closed; include in the results all activities associated with each attraction." We use the native features and syntax of each query language to the best of our knowledge and ability.

*5.1.1 SPARQL.* We begin by considering Complex Patterns in SPARQL. On lines 2-3, we observe a basic graph pattern that finds all attractions having exhibits or rides, looking up such along with the attraction's type. On line 4, the types are filtered down to museums and theme parks. Continuing, we see on line 5 a requirement that the attraction must have a name. Finally, we see on lines 6-8 the requirement that the attractions must not be known to be permanently closed.

Although the individual pieces of the query seem quite intuitive, the overall look of the query strikes us as somewhat unintuitive. We know we are querying a graph, but that does not seem entirely apparent from looking at the query. In our opinion, the SQL-like syntax obscures the "graphy" nature of what we are trying to achieve.

### Complex Patterns in SPARQL

```
1  SELECT ?attraction ?activity {
2    ?attraction a ?type .
3    ?attraction <exhibit>|<ride> ?activity .
4    FILTER(?type IN (<museum>, <theme_park>))
5    FILTER EXISTS { ?attraction <name> ?name }
6    FILTER NOT EXISTS {
7      ?attraction <permanently_closed> true .
8    }
9  }
```

*5.1.2 Cypher.* Moving on to Complex Patterns in Cypher, we take note that the data model for Cypher is property graphs [3]. Thus, we attempt to utilize the feature of "properties" to achieve a simpler query.[14] On line 1, we see a `MATCH` pattern. It quite clearly demonstrates that we are matching edges labelled either `Exhibit` or `Ride`. Then in the where clause on lines 2-6, we apply additional filters taking advantage of node labels and properties. On line 3, we say the attractions must be either museums or theme parks. On line 4, we say the attractions must have names (modeled here as a list-valued property). On lines 5-6, we say that either the `permanently_closed` property must be missing or set to false. The returned values are the `id` property values for the `attraction` and `activity` nodes, as shown on line 7.

---

[11] Meant here as literal "path queries" [3], not to be confused with the name of our language, *PathQuery*.

[12] A measurable and objective analysis of query languages is a difficult feat to achieve and is well beyond the scope of this paper. However, we would be remiss if we did not attempt to compare and contrast with the state of the art in some fashion. We include drawbacks of PathQuery later in the paper as well.

[13] Actually referred to as "complex graph patterns," but we feel that the "graph" part is somewhat misleading, since in our view, some of the language features admitted as "complex graph patterns" are hardly patterns resembling graphs.

[14] We did this by using node labels for types and node properties for associations with literal values like booleans and strings. Additionally, to be somewhat on par with the edge-labelled model, we assumed each node to have an `id` property containing a unique identifier for that node.

**Table 2: Comparison of common language features for SPARQL, Cypher, Gremlin, and PathQuery**

| Language | Data Model | Pattern Composition | Path Queries[11] | Graph Updates |
|---|---|---|---|---|
| SPARQL | edge-labelled graph (RDF) | SQL-ish keywords | regular, plus negated properties | Yes |
| Cypher | property graph | SQL-ish keywords | regular | Yes |
| Gremlin | property graph | path structures | regular | Yes |
| PathQuery | edge-labelled graph | path structures | bounded | No |

Similar to SPARQL, the individual pieces of the query seem quite intuitive, yet the overall look of the query does not particularly resemble a graph. While the `MATCH` statement is very "graphy," the remainder resembles SQL. To be fair, though, this SQL-ish part is not completely unwelcome, because the data model is property graphs, where an SQL-like approach conceivably seems quite appropriate.

**Complex Patterns in Cypher**

```
1 MATCH (attraction)-[:Exhibit|Ride]->(activity)
2 WHERE
3   (attraction:Museum OR attraction:ThemePark)
4   AND EXISTS(attraction.name)
5   AND (NOT EXISTS(attraction.permanently_closed)
6       OR NOT(attraction.permanently_closed))
7 RETURN attraction.id, activity.id
```

*5.1.3 Gremlin.* Turning to Complex Patterns in Gremlin, we find a more navigational syntax. Gremlin also operates on property graphs, so we try to take advantage of that as well. The query begins with `g.V()` which indicates iteration over all nodes in the graph. Lines 1-2 require that the nodes be labelled as either museum or theme park. Line 3 requires that nodes have values for the name property. Lines 4-5 require that nodes have the `permanently_closed` property either missing or set to false. On line 7, we traverse across outgoing edges labelled `exhibit` or `ride` to nodes. We store references to attractions and their corresponding activities on lines 6 and 8. Finally, on line 10, we retrieve the values of the `id` property on those nodes.

This form of querying feels much more "graphy" to us. However, when considering how to implement a system for this language, it is unclear how we could implement flexible and automatically-optimizing query planning for this language.

**Complex Patterns in Gremlin**

```
1 g.V().or(hasLabel('museum'),
2          hasLabel('themePark'))
3   .has('name')
4   .or(has('permanently_closed', false),
5       hasNot('permanently_closed'))
6   .as('attraction')
```

```
7   .out('exhibit', 'ride')
8   .as('activity')
9   .select('attraction', 'activity')
10  .by('id')
```

*5.1.4 PathQuery.* The comparable query in PathQuery is shown in Complex Patterns in PathQuery. Since variations of this example are in previous sections, we leave it as an exercise for the reader to interpret this query. Note that, as in the Sample Output for attractions_v1.pq, the output of this query will map roots, which are attraction `Id`s, to activity `Id`s.

**Complex Patterns in PathQuery**

```
1 @entities
2   .[/type == (Id('/museum'), Id('/theme_park'))]
3   .require(/name)
4   .prohibit(/permanently_closed.[?cur])
5   .(/exhibit, /ride)
```

## 5.2 Navigation

Another dimension of comparison in [3] is the ability to perform navigational queries across a variable number of edges. In this section, we look at a very simple example in each language. The example query is to find the biological class of cat (which is mammal).

*5.2.1 SPARQL.* In Graph Navigation in SPARQL, we see two triple patterns. The first utilizes the "property paths" feature of SPARQL in the predicate position: `<higher_classification>*`. This means that zero or more outgoing edges labelled `<higher_classification>` should be traversed, and each reachable node should be bound to `?classification`. The second triple pattern, however, restricts it down to the one that corresponds to the `<z/class>` classification.

**Graph Navigation in SPARQL**

```
1 SELECT ?classification {
2   <z/cat>
3     <higher_classification>*
4     ?classification .
5   ?classification <rank> <z/class> .
6 }
```

*5.2.2 Cypher.* In Graph Navigation in Cypher, we see an intuitive `MATCH` pattern for finding the classifications of cat. The asterisk in `-[higherClassification*]->` indicates that zero or more outgoing edges labelled `higherClassification` should be traversed until we arrive at a node with an outgoing edge labelled `rank` that points at a node having `/z/class` as its `id` property.

**Graph Navigation in Cypher**

```
1  MATCH ({id: '/z/cat'})
2      -[:higherClassification*]->
3      (classification)
4      -[:rank]->({id: '/z/class'})
5  RETURN classification.id
```

*5.2.3 Gremlin.* In Graph Navigation in Gremlin, we see that we start with every node in the graph, then restrict it to those whose `id` property includes the value `/z/cat`. The `choose` operation allows us to branch based on whether the `/z/cat` node already has an outoing `rank` edge pointing at a node with `id` `/z/class`. If it does, we simply output the `id` value `/z/cat`. Otherwise, we use `repeat` to traverse one or more `higherClassification` edges as necessary until reaching a node with an outgoing edge labelled `rank` pointing at a node with `id` property equal to `/z/class`, and output the `id` property values for those nodes.

**Graph Navigation in Gremlin**

```
1  g.V().has('id', '/z/cat')
2      .choose(
3          values('rank').is(eq('/z/class')),
4          __.values('id'),
5          __.repeat(out('higherClassification'))
6          .until(out('rank').has('id', '/z/class'))
7          .values('id'))
```

*5.2.4 PathQuery.* Whereas SPARQL and Cypher rely on edge quantifiers like `*`, and whereas Gremlin relies on additional path constructs, PathQuery relies on its ability to define functions in order to provide variable-length traversal. Bounded recursion was already discussed in Section 3.3.3, so we will not review it again here.

A drawback of PathQuery's approach is that it does not support unbounded variable-length edge traversals. A maximum bound must always be specified. However, because this feature builds on a general concept of recursive functions, the nature of these "traversals" is much more powerful, since a function definition can be any conceivable PathQuery. In this regard, PathQuery is more similar in power to Gremlin in that the "traversals" are not limited to *regular path queries*.

It should also be pointed out that the other three languages support bounded edge traversals as well.

**Graph Navigation in PathQuery**

```
1  def base Classifications() {
2      [/rank == Id('/z/class')]
3  }
4
5  def recur<10> Classifications() {
6      (
7          [/rank == Id('/z/class')],
```

```
8          /higher_classification.Classifications()
9      )
10 }
11
12 Id('/z/cat').Classifications()
```

### 5.3 Notably Missing Features

It is worth calling out that PathQuery is missing some notable features that are included in some of the other languages we considered.

- We already mentioned the lack of support for *regular path queries* in Section 5.2.4.
- PathQuery does not currently have the ability to traverse edges regardless of predicate, although we expect to add such a "wildcard predicate" in the near future.
- PathQuery does not have a way to perform aggregations based on groupings of arbitrary variable values. Put more simply, there is no generic `GROUP BY` feature. Aggregates in PathQuery are always grouped by input.
- PathQuery does not have support for graph creation or updates.
- As mentioned in Section 3.1.3, PathQuery runs over a single graph. It does not have a mechanism for accessing more than a single graph in a query (e.g., unlike SPARQL which has "named graphs").
- Although PathQuery includes a language feature for inspecting triples accessed during the query (not discussed elsewhere herein), its behavior is implemented-dependent, and is generally insufficient for supporting the entire property graph model.

Note that these features are missing from the language so far because we have yet to need them (the wildcard predicate excepted). PathQuery language feature development has so far been driven primarily by the needs of internal clients.

## 6 CONCLUSIONS

In this paper, we introduced PathQuery, Google's graph query language, which is used for serving the Knowledge Graph. It is the interface to the Knowledge Graph for major services and products like Search, Maps, and Assistant, and it is designed to be accessible to generalist developers. PathQuery has withstood tests of time and scale at Google, and we also believe it manages to be declarative without imposing relational constructs on the language. Ultimately, this work is intended to share our language, insight, and experience with the graph query language and database community.

We described the language, first building an extended example driven by an example use case, then enumerating and describing individual language features. We also compared and contrasted PathQuery with three other graph query languages – SPARQL, Cypher, and Gremlin – describing the intuition for why we passed over these options and created a new query language. We also highlighted features missing from PathQuery that are present in those three other languages.

## ACKNOWLEDGMENTS

We would like to recognize as original contributors to the language: Warren Harris (posthumous), Eduardo Ribas, and Rene Zhang. We would also like to recognize the following as significant contributors in developing PathQuery systems internally: Bei Li, Murilo Vasconcelos, Mandi Wang, Peter Tran, and Ayush Shah.